\documentclass[letterpaper,10pt,conference]{ieeeconf}

\IEEEoverridecommandlockouts                              % This command is only needed if 
\usepackage{amsmath,amssymb}
\usepackage{mathtools}
\usepackage{array}
\usepackage{multirow}
\usepackage{booktabs}
\usepackage{hyperref}
\usepackage[compress,noadjust]{cite}
\usepackage{xcolor}
\usepackage[linesnumbered,ruled,vlined]{algorithm2e}

\title{\LARGE\bf%
Data-driven Koopman mode approximation:\\A neural power iteration algorithm
}

\author{Guillaume O.~Berger$^{1}$, and Rapha\"el M.~Jungers$^{1}$% <-this % stops a space
\thanks{$^{1}$Both authors are with ICTEAM, UCLouvain, Louvain-la-Neuve, Belgium.
GB is an FNRS Postdoctoral Researcher.
RJ is an FNRS honorary Research Associate.
This project has received funding from the European Research Council (ERC) under the European Union's Horizon 2020 research and innovation programme under grant agreement no.~864017 (L2C), from the Horizon Europe programme under grant agreement no.~101177842 (Unimaas), and from the ARC (French Community of Belgium) (SIDDARTA).
Emails: {\{guillaume.berger,raphael.jungers\}@uclouvain.be}.}%
}

\newcommand{\C}{\mathbb{C}}
\newcommand{\R}{\mathbb{R}}
\newcommand{\N}{\mathbb{N}}

\DeclareMathOperator*{\argmin}{arg\,min}
\DeclareMathOperator{\spann}{span}

\newcommand{\calD}{\mathcal{D}}
\newcommand{\calF}{\mathcal{F}}
\newcommand{\calK}{\mathcal{K}}
\newcommand{\calV}{\mathcal{V}}
\newcommand{\calX}{\mathcal{X}}

\newtheorem{definition}{Definition}
\newtheorem{theorem}{Theorem}

\newtheorem{proposition}{Proposition}
\newtheorem{corollary}{Corollary}

\newtheorem{remark}{Remark}

\colorlet{mylinkcolor}{red!80!black}
\colorlet{myurlcolor}{green!50!black}
\colorlet{mysectioncolor}{blue!50!black}
\hypersetup{%
    colorlinks=true,
    linkcolor=mylinkcolor,
    urlcolor=myurlcolor,
    hypertexnames%
}

\SetCommentSty{mycommfont}

\newif\ifextended
\extendedtrue

\begin{document}

\maketitle
\thispagestyle{empty}
\pagestyle{empty}

%%%%%%%%%%%%%%%%%%%%%%%%%%%%%%%%%%%%%%%%%%%%%%%%%%%%%%%%%%%%%%%%%%%%%%%%%%%%%%%%
\begin{abstract}
This paper proposes a novel data-driven algorithm to approximate the dominant eigenfunctions (aka.~modes) of the Koopman operator of nonlinear dynamical systems using neural networks.
The relevance of learning the dominant Koopman modes is to approximate nonlinear dynamics by linear ones in a lifted space, thereby enabling simplified control and analysis.
To fight the curse of dimensionality arising from using expressive templates (here neural networks) for the mode approximation, the proposed method leverages a power-iteration scheme that directly learns the dominant Koopman modes without explicitly constructing the projection of the Koopman operator on the template of functions.
Our approach connects to other approaches in the literature that avoid the curse of dimensionality by learning small dictionaries of functions, but differs from them in that we do not require ``anti-collapse mechanisms'' to ensure that the learned dictionary is expressive enough to approximate the Koopman operator since our power-iteration scheme is designed to converge toward the dominant modes of the projected Koopman operator.
The approach is fully data-driven, requiring only sampled state transitions.
Theoretical guarantees are provided, showing convergence under increasing sample size and network width (in connection with the neural tangent kernel theorem).
Numerical experiments demonstrate that the method achieves accurate and smooth approximations of dominant modes while avoiding the limitations of traditional techniques such as extended dynamic mode decomposition.
\end{abstract}

%%%%%%%%%%%%%%%%%%%%%%%%%%%%%%%%%%%%%%%%%%%%%%%%%%%%%%%%%%%%%%%%%%%%%%%%%%%%%%%%
\section{Introduction}

Koopman theory~\cite{koopman1931hamiltonian} offers a powerful alternative to state-based techniques for studying nonlinear dynamical systems
\[
x^+ = F(x).
\]
Rather than studying the evolution of state variables $x\in\R^n$, the Koopman approach studies the evolution of observables, i.e., functions $f:\R^n\to\C$, under the action of the system.
The overarching idea of this approach is that the action of the system on the observables can be represented by a linear operator, called \emph{Koopman operator}, defined by
\[
\calK:\calF\to\calF, \quad (\calK f)(x) = f(F(x)),
\]
where $\calF$ is a suitable set of observables.
This allows to apply techniques from linear operator theory, such as spectral analysis, to study the system, including convergence analysis~\cite{mauroy2016global}, chaos~\cite{mezic2020spectrum}, model order reduction~\cite{mezic2005spectral}, and control~\cite{kaiser2021datadriven}; see~\cite{budivsic2012applied,otto2021koopman,brunton2022modern,colbrook2025anintroductory} for recent surveys.

Another key advantage of the Koopman approach is that it is easily applicable in the data-driven setting.
Indeed, the action of the system on an observable $f$ can be approximated by its action on a subset of sample states $\{x_i\}_{i=1}^N\subseteq\R^n$, which gives a straightforward data-driven approximation
\[
(\calK f)(x_i^{})=f(x_i^+), \quad i=1,\ldots,N,
\]
where $\{(x_i^{},x_i^+)\}_{i=1}^N$ are sample one-step trajectories of the system.

However, the set $\calF$ of observables being typically infinite-dimensional, Koopman-based techniques usually require projection on a finite-dimensional subspace~\cite{colbrook2024themultiverse}.
More precisely, given a finitely-generated subspace $\calD\subseteq\calF$, one studies the projected Koopman operator
\[
K_\calD:\calD\to\calD, \quad K_\calD=\Pi_\calD\circ\calK,
\]
where $\Pi_\calD$ is a projection on $\calD$.
The projection introduces approximation errors that need to be taken into account in the analysis of the system through the projected Koopman operator.
This raises the fundamental question of the selection of the projection subspace $\calD$, which needs to balance computational efficiency and accuracy of the approximation~\cite{colbrook2024rigorous}.

The dynamic mode decomposition (DMD)~\cite{schmid2010dynamic} approach uses the set of linear observables as space $\calD$.
The extended DMD (EDMD)~\cite{williams2015adatadriven} extends the former with a user-provided dictionary of functions (e.g., polynomials of degree $\leq d$) as basis of $\calD$.
The Hankel DMD (aka.~time-delay DMD)~\cite{arbabi2017ergodic} is a particular case of EDMD where the dictionary is built from a user-provided function $g$ on which the Koopman operator is applied $m$ times.

The reliance on a user-provided dictionary is a significant limitation because good approximations of the Koopman operator typically require expressive dictionaries.
For instance, bistable systems have discontinuous dominant eigenmodes of their Koopman operator~\cite{mauroy2016global}; thereby requiring polynomials of high degree for good approximations of such modes.

Consequently, in recent years, there has been significant research effort devoted to exploiting the expressive power of (deep) neural networks to parametrize observables~\cite{li2017extended,takeishi2017learning,lusch2018deep,alfordlago2022deep,mondal2023efficient}.
The main challenge with this approach is that there is no clear basis of the associated space $\calD$.
The learning problem is therefore formulated as the one of finding a dictionary of neural network observables along with the associated projected Koopman operator:
\begin{equation}\label{eq:neural-koopman-naive}
\min_{\substack{\theta_i\in\R^{n_\theta}\\K\in\C^{m\times m}}} \, \lVert [\calK\phi^{\theta_1},\ldots,\calK\phi^{\theta_m}] - [\phi^{\theta_1},\ldots,\phi^{\theta_m}]K \rVert,
\end{equation}
where $\phi^\theta\in\calF$ is a neural network parametrized by $\theta\in\R^{n_\theta}$, and $K$ is the matrix representation of the projected Koopman operator associated to $\calD=\spann\{\phi^{\theta_1},\ldots,\phi^{\theta_m}\}$.
Although attractive for its simplicity, formulation~\eqref{eq:neural-koopman-naive} is problematic because of the \emph{collapse problem}: generically, the dictionary collapses to a trivial dictionary with only a few linearly independent observables (e.g., $\phi^{\theta_i}=0$ for all $i$)~\cite{takeishi2017learning,li2017extended}.

To prevent collapse, state-of-the-art approaches (see refs above) require that the state can be (approximately) retrieved from the encoding $[\phi^{\theta_1}(x),\ldots,\phi^{\theta_m}(x)]$.
This is enforced either explicitly by fixing some of the observables $\phi^{\theta_i}$ so that $x$ can be retrieved from those observables~\cite{li2017extended}, or implicitly by adding a penalty term in the loss of~\eqref{eq:neural-koopman-naive}:
\[
\rho \lVert \mathrm{Id} - \Delta^{\theta_0}\circ[\phi^{\theta_1},\ldots,\phi^{\theta_m}] \rVert,
\]
where $\mathrm{Id}$ is the identity function and $\Delta^\theta:\C^m\to\R^n$ is a trainable neural network playing the role of the \emph{decoder}~\cite{takeishi2017learning,lusch2018deep,alfordlago2022deep,mondal2023efficient}.\footnote{In other words, we impose that $x\approx\Delta^{\theta_0}(\phi^{\theta_1}(x),\ldots,\phi^{\theta_m}(x))$.}
The drawback of these state-retrieval approaches is that they require a number of observables that grows with the dimension of the system, thereby limiting their use in systems of large dimension.
Furthermore, the decoder-based approach requires to train an additional network, which makes training slower and more data-sensitive.

\subsection*{Contributions}

In this work, we propose a novel iterative method, based on power iterations, to learn the dominant modes of the projected Koopman operator.
A \emph{mode} of the projected Koopman operator is an eigenfunction of it, that is, a non-zero observable $g\in\calD$ such that $K_\calD g=\lambda g$ for some $\lambda\in\C$ called the associated eigenvalue.
A mode is \emph{dominant} if its eigenvalue is large in modulus compared to the other eigenvalues.
Power iteration is a numerical iterative algorithm that allows one to approximate the $m$ dominant modes of a linear operator, where $m$ is a user-provided parameter~\cite{golub2013matrix}.

Conceptually, our approach works as follows: starting from a set of observables $\{f_1,\ldots,f_m\}$, we apply recursively $K_\calD$ to each observable $f_i$.
Under appropriate assumptions (e.g., separation of the dominant $m$ eigenvalues of $K_\calD$~\cite{golub2013matrix}), the set $\{K_\calD^k f_1,\ldots,K_\calD^k f_m\}$ converges toward a basis of the subspace spanned by the dominant $m$ modes of $K_\calD$.
To avoid ill-conditioning, a normalization step can be applied periodically.
The above conceptual approach is made practical by (i) using data to approximate the action of the exact Koopman operator, and (ii) using gradient descent to project back on the set of neural-network observables.

Concretely, the contributions of this work are as follows:
\begin{itemize}
    \item {\bf Theory:} We provide the first power-iteration-based algorithm to approximate the dominant modes of the projected Koopman operator, without requiring its explicit construction; thereby alleviating the need of large user-provided dictionaries, or anti-collapse mechanisms for learned dictionaries;\\
    We discuss the soundness of our approach and of its data-driven neural-network-based implementation (see below), based on the neural tangent kernel theorem~\cite{jacot2018neural};
    \item {\bf Implementation:} We provide a tractable data-driven implementation of the above algorithm for observables parametrized by neural networks;
    \item {\bf Experiments:} We demonstrate the usefulness of our approach on numerical examples.
\end{itemize}

\subsection*{Applications of Koopman mode decomposition}

The Koopman mode decomposition (KMD), introduced by Mezi\'c~\cite{mezic2005spectral}, exploits the spectral properties of the Koopman operator to decompose observables of nonlinear dynamical systems into components with simple linear evolution; see also~\cite{budivsic2012applied,mauroy2016global,bevanda2021koopman} for surveys.
Concretely, if $f=\sum_{i=1}^\infty \alpha_i g_i$, where $g_i$ are eigenfunctions of $\calK$ with eigenvalues $\lambda_i$, then $\calK^k f = \sum_{i=1}^\infty \alpha_i^{} \lambda_i^k g_i^{}$.
Such an eigenfunction expansion is not available for every dynamical system or every observable space: the point spectrum may be insufficient or even absent, and Koopman eigenfunctions need not form a basis~\cite{mezic2020spectrum}.
The KMD, if it exists, however shows that only the dominant modes will dictate the long-term dynamic (i.e., the value of $\calK^kf$ for large $k$) because $\lambda_i^k$ will converge faster to zero for the dominated modes than for the dominant ones.
Notably, the modes with eigenvalue one give the basins of attraction of the system, while the sub-unitary dominant eigenvalues give the rate of convergence of the system toward attractors~\cite{mauroy2016global,bevanda2021koopman}; see also Section~\ref{sec:problem}.

It is also well known that the projection of the Koopman operator on a subspace introduces perturbations in the spectrum of $\calK$.
This needs to be taken into account in the analysis of the system using the projected Koopman operator; see, e.g.,~\cite{colbrook2024rigorous,colbrook2024themultiverse}.
At the level of $\calK$ itself, Generalized Laplace Analysis (GLA)~\cite{mezic2005spectral} provides a related spectral-extraction viewpoint based on weighted time averages; notably, the averaging can separate eigenvalues of equal modulus but different phases, which ordinary power iteration cannot.
Here, however, our goal is to learn the dominant modes of the projected operator $K_\calD$ rather than analyze the spectrum of $\calK$ itself.
We therefore leave these spectral-approximation questions aside, while noting that they motivate the use of expressive sets $\calD$.

Control can also be achieved with Koopman mode decomposition when applied to the Koopman operator of open-loop systems~\cite{kaiser2021datadriven,otto2021koopman}.
This will be left for future work as well.

\subsection*{Other related work}

Recently, several research efforts in the field of machine learning and artificial intelligence have been devoted to learning the dynamics of ``systems'' or environments in reduced-order latent space representations allowing for prediction, analysis, and planning~\cite{ha2018world,assran2023selfsupervised,ulmen2025learning}.
For instance, the joint embedded predictive architecture (JEPA) learns together a latent representation of the state of the environment using neural network encoders as well as a neural network predictive model for the evolution of the latent representation over time.
These approaches connect to the philosophy of this work in that they also aim to learn simultaneously the latent representation of the state and the predictive model in the latent space.
Consequently, like us, they face the same risk of collapse of their model during training.
The differences with our approach are twofold: (i) they use neural network as predictive models, whereas we use linear models (encoded by the matrix $K_\calD$); and (ii) they fight collapse with techniques different from our power-iteration-based approach (e.g., contrastive samples, entropy maximization, decoders, etc.)~\cite{terver2026alightweight}.

\subsection*{Notation}

Throughout this paper, we fix $n\in\N_{>0}$ and a \emph{state space} $\calX\subseteq\R^n$.
Let $(\calX,\Sigma,\mu)$ be a probability space on $\calX$, with $\sigma$-algebra $\Sigma$ and probability measure $\mu$.
We let $\calF$ be the space $L^2(\calX,\Sigma,\mu)$ endowed with the norm $\lVert\cdot\rVert$.\footnote{That is, $\calF$ consists of the $\Sigma$-measurable functions $f:\calX\to\C$ such that $\lVert f\rVert^2\triangleq\mathbb{E}_\mu[\lvert f\rvert^2]<\infty$.}
Note that $\calF$ is a Hilbert space with inner product $\langle f,g\rangle\triangleq\mathbb{E}_\mu[f\bar{g}]$~\cite{friedman1982foundations}.
We denote by $\mathrm{i}$ the imaginary unit.

A \emph{feed-forward neural network} from $\R^n$ to $\C$ is generically denoted by $\phi^\theta$, where $\theta\in\R^{n_\theta}$ is a parameter vector for the weights and biases of the network.
Concretely, for a number $L\in\N_{>0}$ of hidden layers with respective widths $c_1,\ldots,c_L\in\N_{>0}$ and an activation function $\sigma:\R\to\R$, the network $\phi^\theta$ is defined by
\begin{align*}
y_0 &= x, \\
y_{k+1} &=\sigma(W^\theta_k y_k + b^\theta_k), \quad k=0,\ldots,L-1, \\
\phi^\theta(x) &= [1,\mathrm{i}](W^\theta_L y_L + b^\theta_L),
\end{align*}
where the weight matrices $W^\theta_k\in\R^{c_{k+1}\times c_k}$ and bias vectors $b^\theta_k\in\R^{c_{k+1}}$ (with $c_0=n$ and $c_{L+1}=2$) are parametrized by $\theta$, and $\sigma$ is applied \emph{entrywise} on vectors.

%%%%%%%%%%%%%%%%%%%%%%%%%%%%%%%%%%%%%%%%%%%%%%%%%%%%%%%%%%%%%%%%%%%%%%%%%%%%%%%%
\section{Problem statement}\label{sec:problem}

We consider a discrete-time dynamical system of the form
\[
x(t+1) = F(x(t)), \quad t\in\N,
\]
where $F:\calX\to\calX$ is a continuous map.
In particular, $\calX$ is assumed to be forward invariant under $F$, so that $F(x)\in\calX$ for every $x\in\calX$.
We assume throughout that $f\circ F\in\calF$ for every $f\in\calF$, so that the composition operator induced by $F$ maps $\calF$ into itself.

Next, we define the Koopman operator of this system.

\begin{definition}
Let $F:\calX\to\calX$ be a continuous map.
The \emph{Koopman operator} of $F$ is the linear operator $\calK:\calF\to\calF$ defined by
\[
\calK f = f\circ F, \quad \forall\,f\in\calF,
\]
i.e., $(\calK f)(x) = f(F(x))$ for all $x\in\calX$ and $f\in\calF$.
\end{definition}

Eigenvalues and eigenfunctions of this operator are defined in the traditional way:

\begin{definition}
Let $\calK:\calF\to\calF$ be a linear operator.
A non-zero function $g\in\calF$ is an \emph{eigenfunction} of $\calK$ if there exists $\lambda\in\C$, called an \emph{eigenvalue}, such that $\calK g=\lambda g$.
\end{definition}

The connections between the eigenvalues and eigenfunctions of the Koopman operator and the properties of the dynamical system (such stability, chaos, and ergodicity) have been extensively studied~\cite{mezic2020spectrum,mauroy2016global}.
We mention below two results that we will use in the numerical experiments.

\begin{proposition}[{\cite[\S3.A]{mauroy2016global}}]
Let $F:\calX\to\calX$ be a continuous map and $\calK$ its Koopman operator.
Let $x_*\in\calX$ be an asymptotically stable equilibrium point of $F$.
Let $g\in\calF$ be an eigenfunction of $\calK$ with eigenvalue $1$.
Assume that $g$ is continuous at $x_*$.
Then, $g$ is constant on the basin of attraction of $x_*$.
\end{proposition}

\begin{proposition}[{\cite[\S3.A]{mauroy2016global}}]
Let $F:\calX\to\calX$ be a continuous map and $\calK$ its Koopman operator.
Let $x_*\in\calX$ be a globally asymptotically stable equilibrium point of $F$.\footnote{The assumption of \emph{global} attractivity can be made without loss of generality since it suffices to restrict $\calX$ to the basin of attraction of $x_*$.}
Define the set $\calF_0=\{f\in\calF:f(x_*)=0\}$.
It holds that:
\begin{itemize}
    \item[i)] $\calF_0$ is invariant for $\calK$, i.e., $\calK\calF_0\subseteq\calF_0$.
\end{itemize}
Let $g\in\calF_0$ be an eigenfunction of $\calK$ with eigenvalue $\lambda\in\C$.
Assume that $g$ is continuous at $x_0$.
It holds that:
\begin{itemize}
    \item[ii)] $\lvert\lambda\rvert<1$.
    \item[iii)] $x\mapsto\lvert g(x)\rvert$ provides a \emph{semi-Lyapunov function} (``semi'' in the sense that it may be $0$ at $x\neq x_*$) for $F$.
\end{itemize}\vskip0pt
\end{proposition}

\subsection*{Projected Koopman operator}

As mentioned earlier, working in an infinite-dimensional space of observables is typically not tractable computationally.
Therefore, we usually restrict our attention to a finite-dimensional subspace of observables, denoted $\calD\subseteq\calF$.
Since it is not guaranteed that $\calD$ is invariant for $\calK$, we need to project back on $\calD$ after application of $\calK$.
Given a closed linear subspace $\calD\subseteq\calF$, the \emph{projection} on $\calD$ is the linear operator $\Pi_\calD:\calF\to\calF$ defined by
\begin{equation}\label{eq:projection}
\Pi_\calD f = \argmin_{g\in\calD}\,\lVert f-g\rVert.
\end{equation}
The \emph{projected Koopman operator} on $\calD$ is then defined as
\[
K_\calD:\calD\to\calD, \quad K_\calD=\Pi_\calD\circ\calK.
\]

In the following, we assume that $\calD$ is a finite-dimensional linear subspace of the Hilbert space $\calF$, and we denote its dimension by $D$ (typically, $D\gg1$).
Every finite-dimensional linear subspace of a normed space is closed; hence $\calD$ is closed and the orthogonal projection $\Pi_\calD$ in~\eqref{eq:projection} is well defined and linear.
We also assume that $\calK\calD\subseteq\calF$.\footnote{This is the case for instance if $\calD$ contains only piecewise continuous functions and $F$ is a continuous map.}
With these assumptions, $K_\calD$ is well defined and is a finite-dimensional linear operator.
We denote its eigenvalues (counted with algebraic multiplicity) by $\lambda_1,\ldots,\lambda_D\in\C$, ordered as
\begin{equation}\label{eq:eigenvalues}
\lvert\lambda_1\rvert\geq\lvert\lambda_2\rvert\geq\cdots\geq\lvert\lambda_D\rvert.
\end{equation}

The goal of this paper is to approximate the $m$ dominant eigenvalues and associated invariant subspace of the \emph{projected} operator $K_\calD$, for some $1\leq m\ll D$, without computing $K_\calD$ explicitly.
Thus, the primary object targeted by our power iteration is $K_\calD=\Pi_\calD\circ\calK$, rather than the infinite-dimensional Koopman operator $\calK$ itself (the relation between eigenfunctions of $K_\calD$ and those of $\calK$ is a separate approximation issue and can be affected by spectral pollution, as recalled in Remark~\ref{rem:pollution} below).
We introduce in the next section a power-iteration-based algorithm that does not require an explicit matrix representation of $K_\calD$ (nor an explicit basis of $\calD$, e.g., when the projection is implemented through neural regression).
Furthermore, this approach can be implemented in a fully data-driven way.

\begin{remark}\label{rem:pollution}
As mentioned in the introduction, the projection of the Koopman operator may induce \emph{spectral pollution}, i.e., eigenfunctions of $K_\calD$ may not be good approximations of eigenfunctions of $\calK$~\cite{colbrook2024rigorous}.
We stress that this is an issue common to all Koopman methods based on projection on a finite-dimensional subspace~\cite{schmid2010dynamic,williams2015adatadriven,arbabi2017ergodic,takeishi2017learning,lusch2018deep,alfordlago2022deep,mondal2023efficient,colbrook2024themultiverse}.
Post-processing techniques can be used to filter out spurious eigenfunctions~\cite{colbrook2024rigorous}; we will use these techniques in our numerical experiments (Section~\ref{sec:experiments}) to compare the accuracy of different methods.
\end{remark}

%%%%%%%%%%%%%%%%%%%%%%%%%%%%%%%%%%%%%%%%%%%%%%%%%%%%%%%%%%%%%%%%%%%%%%%%%%%%%%%%
\section{Power-iteration-based algorithm}

We present and analyze our power-iteration-based algorithm to compute the dominant eigenvalues of $K_\calD$, which constitutes the main theoretical contribution of this work.
We first present its theoretical \emph{idealized} version (which is rather straightforward), then a practical neural data-driven version.
We discuss the soundness of the latter, based on the neural tangent kernel theorem.

\subsection{Idealized algorithm}

This idealized version is not applicable in most practical situations, but nevertheless conveys the main theoretical ideas behind the algorithm.
The base idea is to simply apply the power iteration algorithm on the operator $K_\calD$.
The power iteration algorithm consists in choosing an initial set of linearly independent functions $\{f_1^{(0)},\ldots,f_m^{(0)}\}\subseteq\calD$, and applying recursively $K_\calD$ to each of these functions.
Possibly, a normalization step can be applied regularly to avoid ill-conditioning.
A pseudo-code is presented in Algorithm~\ref{algo:idealized}.

\begin{algorithm}
\caption{Idealized Koopman power iteration}
\label{algo:idealized}
\DontPrintSemicolon
\KwData{Koopman operator $\calK$, subspace $\calD\subseteq\calF$ and projection $\Pi_\calD$, \# modes $m\in\N_{>0}$}
$\{f_j^{(0)}\}_{j=1}^m\gets\text{select $m$ lin.~ind.~functions from $\calD$}$ \label{line:initial}\;
\For{$k=0,1,2,\ldots$}{
    $\{h_j^{(k+1)}\}_{j=1}^m\gets\{\calK f_j^{(k)}\}_{j=1}^m$\;
    $\{g_j^{(k+1)}\}_{j=1}^m\gets\{\Pi_\calD h_j^{(k+1)}\}_{j=1}^m$\;
    $\{f_j^{(k+1)}\}_{j=1}^m\gets\mathrm{OrthoBasis}(\{g_j^{(k+1)}\}_{j=1}^m)$ \tcp{Extract orthonormal basis if wanted}
}
\end{algorithm}

Classical results from numerical linear algebra guarantees convergence of the power iteration algorithm toward the dominant modes under assumption of separation of the modulus of the $m$th and $(m+1)$st eigenvalues.

\begin{proposition}[{\cite[Th.~7.3.1]{golub2013matrix}}]
Consider a Koopman operator $\calK$ and a finite-dimensional subspace $\calD\subseteq\calF$.
Let the eigenvalues of $K_\calD$ be ordered as in~\eqref{eq:eigenvalues}.
Assume that $\lvert\lambda_m\rvert>\lvert\lambda_{m+1}\rvert$.
Let $\calV_m\subseteq\calD$ be the invariant subspace of $K_\calD$ associated to the first $m$ eigenvalues.
Consider an execution of Algorithm~\ref{algo:idealized}.
Then, for almost all\footnote{With respect to the natural Lebesgue measure on $\calD^m$, which means the Lebesgue measure than can be derived from any orthonormal basis of $\calD$.} choices of $\{f_1^{(0)},\ldots,f_m^{(0)}\}\subseteq\calD$ in line~\ref{line:initial}, $\spann\{f_1^{(k)},\ldots,f_m^{(k)}\}$ converges towards $\calV_m$ when $k\to\infty$.
\end{proposition}

\subsection{Neural data-driven algorithm}

The purpose of the neural data-driven version is to relax the computation of $\calK f_j$ and $\Pi_\calD h_j$ in Algorithm~\ref{algo:idealized}, which are both intractable for most practical problems.\footnote{$\calK f_j$ requires knowing $f_j$ at every $x$.
$\Pi_\calD h_j$ requires either an orthonormal basis + numerical integration or a basis + numerical integration + the inversion of a $D\times D$ Gram matrix~\cite{williams2015adatadriven}.}

Therefore, the projection $\Pi_\calD$ is first replaced by a data-driven approximation.
Given $N$ samples $\{x_i\}_{i=1}^N\subseteq\calX$ drawn i.i.d.~from $\mu$, we approximate $\langle\cdot,\cdot\rangle$ and $\lVert\cdot\rVert$ by Monte--Carlo approximations $\langle\cdot,\cdot\rangle'_N$ and $\lVert\cdot\rVert'_N$:
\begin{equation}\label{eq:mc-norm}
\langle f,g\rangle'_N \triangleq \frac1N \sum_{i=1}^N f(x_i)\bar{g}(x_i), \quad {\lVert f\rVert'_N}^2 \triangleq \langle f,f\rangle'_N.
\end{equation}
This allows to approximate $\Pi_\calD$ by replacing ``$\lVert\cdot\rVert$'' by ``$\lVert\cdot\rVert'_N$'' in~\eqref{eq:projection}.
The resulting operator is denoted by $\Pi'_{N,\calD}$:\footnote{We assume without loss of generality (possibly at the cost of taking more samples) that $\lVert\cdot\rVert'_N$ is a norm on $\calD$, so that the argmin in \eqref{eq:projection-approx} is unique.}
\begin{equation}\label{eq:projection-approx}
\Pi'_{N,\calD} f = \argmin_{g\in\calD}\,\lVert f-g\rVert'_N.
\end{equation}
When the elements of $\calD$ are represented by a neural network $\phi^\theta$, the argmin in~\eqref{eq:projection-approx} reduces to classical function regression using $\ell^2$ loss:
\begin{equation}\label{eq:nn-projection}
\Pi'_{N,\Phi}f = \argmin_{\theta\in\R^{n_\theta}}\,\frac1N \sum_{i=1}^N \left\lvert\phi^\theta(x_i)-f(x_i)\right\rvert^2
\end{equation}
(we argue in Section~\ref{ssec:analysis} that this yields a linear operator under appropriate assumptions).
\eqref{eq:nn-projection} can be solved by using gradient descent, Adam, AdamW, etc.~\cite{loshchilov2017decoupled}.

Following Algorithm~\ref{algo:idealized}, $\Pi'_{N,\Phi}$ must be applied to $h_j^{(k+1)}=\calK f_j^{(k)}$.
Hence, $f(x_i)$ in~\eqref{eq:nn-projection} must be replaced by $h_j^{(k+1)}(x_i)$.
Since $\calK f_j(x_i)=f_j(F(x_i))$, it suffices to compute
\begin{equation}\label{eq:data-driven-successors}
x^+_i\triangleq F(x_i) \quad i=1,\ldots,N,
\end{equation}
and let $h_j^{(k+1)}(x_i)=f_j^{(k)}(x_i^+)$.
In particular, if the current iterate is $f_j^{(k)}=\phi^{\theta_j^{(k)}}$, the next pre-orthogonalization iterate is given by $g_j^{(k+1)}=\phi^{\zeta_j^{(k+1)}}$ where
\begin{equation}\label{eq:nn-iteration-explicit}
\zeta_j^{(k+1)}
\in\argmin_{\theta\in\R^{n_\theta}}
\frac1N\sum_{i=1}^N
\left|\phi^\theta(x_i)-\phi^{\theta_j^{(k)}}(x_i^+)\right|^2.
\end{equation}
\eqref{eq:nn-iteration-explicit} makes explicit that no representation of $\calK$ is constructed: at each power iteration, a neural network is simply regressed on the values of the previous iterate evaluated at the successor samples.
Hence, the computations are fully data-driven, based only on i.i.d.~samples $\{x_i\}_{i=1}^N\subseteq\calX$ and their successors $\{x_i^+\}_{i=1}^N$ by the system as in~\eqref{eq:data-driven-successors}.
A pseudo-code of the neural data-driven algorithm is presented in Algorithm~\ref{algo:data-driven}.

\begin{algorithm}
\caption{Neural data-driven Koopman power iteration}
\label{algo:data-driven}
\DontPrintSemicolon
\KwData{Samples $\{x_i\}_{i=1}^N\subseteq\calX$ and successors $\{x_i^+\}_{i=1}^N$, neural network $\phi^\theta$, \# modes $m\in\N_{>0}$}
$\{\theta_j^{(0)}\}_{j=1}^m\gets\text{initialize $m$ param.~vec.~from $\R^{n_\theta}$}$\; \label{line:nn-initial}
\For{$k=0,1,2,\ldots$}{
    \For{$j=1,\ldots,m$}{
        $\zeta_j^{(k+1)}\gets\text{solve regression~\eqref{eq:nn-iteration-explicit}}$\;
        $g_j^{(k+1)}\gets\phi^{\zeta_j^{(k+1)}}$\;
    }
    $\{\theta_j^{(k+1)}\}_{j=1}^m\gets\mathrm{OrthoBasis}(\{g_j^{(k+1)}\}_{j=1}^m)$
    \label{line:nn-ortho} \tcp{cf.~Remark~\ref{rem:ortho-nn}}
}\end{algorithm}

\begin{remark}\label{rem:ortho-nn}
The orthonormalization of the basis of neural networks (line~\ref{line:nn-ortho} in Algorithm~\ref{algo:data-driven}) can be done easily from the samples as follows:
\begin{itemize}
    \item[1)] At a given iteration $k$, compute the vectors $v_1,\ldots,v_m\in\C^N$ defined by $v_j=[g_j^{(k+1)}(x_i)]_{i=1}^N$.
    \item[2)] Compute an orthonormal basis $\{q_1,\ldots,q_m\}$ of $\{v_1,\ldots\allowbreak,v_m\}$ (e.g., by computing the QR decomposition of the matrix $[v_1,\ldots,v_m]\in\C^{N\times m}$~\cite{golub2013matrix}).
    \item[3)] For all $1\leq j\leq m$, find
    \[
    \theta_j^{(k+1)}\in\argmin_{\theta\in\R^{n_\theta}}\,\frac1N \sum_{i=1}^N \left\lvert\phi^\theta(x_i)-q_j[i]\right\rvert^2.
    \]
\end{itemize}
If the fitting error in 3) is less than or equal to $\epsilon$, it follows that for all $1\leq j_1,j_2\leq m$, $\lvert \langle\phi^{\theta_{j_1}},\phi^{\theta_{j_2}}\rangle'_N - \delta_{j_1,j_2}\rvert\leq2\epsilon+O(\epsilon^2)$.
In other words, $\{\phi^{\theta_1},\ldots,\phi^{\theta_m}\}$ is $\epsilon$-orthonormal with respect to $\langle\cdot,\cdot\rangle'_N$.\footnote{Note that exact orthonormality is not needed since our aim is merely to avoid ill-conditioning.}
Furthermore, under the assumption that the set of functions parametrized by $\phi^\theta$ forms a linear subspace (see Section~\ref{ssec:analysis}), the global minimum of fitting error in 3) achieves $\epsilon=0$.~\hfill$\triangleleft$
\end{remark}

\subsection{Analysis of neural data-driven v.s.~idealized}\label{ssec:analysis}

We analyze the neural data-driven algorithm, in particular its soundness to approximate the idealized algorithm.

First, we start with the Monte--Carlo approximation of the norm and inner product.
By the strong law of large numbers, we have that both approximations are sound as $N\to\infty$:

\begin{theorem}[{\cite[\S~8]{athreya2006measure}}]\label{tmh:LLN}
Let $\{x_i\}_{i=1}^\infty\subseteq\calX$ be a sequence of i.i.d.~samples from $\mu$.
Let $f,g\in\calF$.
With probability $1$ on the sampling of $\{x_i\}_{i=1}^\infty$, it holds that
\[
\langle f,g\rangle'_N\to\langle f,g\rangle \quad \text{and} \quad \lVert f\rVert'_N\to\lVert f\rVert, \quad \text{as} \quad N\to\infty.
\]\vskip0pt
\end{theorem}

As a corollary, we obtain the main result of this section:

\begin{corollary}[Main result]\label{cor:dd-convergence}
Consider a Koopman operator $\calK$ and a neural network $\phi^\theta$.
Assume that $\Pi'_{N,\Phi}$ in~\eqref{eq:nn-projection} is a linear operator with image set denoted by $\calD$ (which is a linear subspace).
Assume that $\calD\subseteq\calF$ and $\calK\calD\subseteq\calF$.
% Assume that~\eqref{eq:nn-projection-koopman} always finds the global minimum.
Let $\{x_i\}_{i=1}^\infty\subseteq\calX$ be a sequence of i.i.d.~samples from $\mu$.
\begin{itemize}
    \item[i)] For any $f\in\calF$, it holds that $\Pi'_{N,\Phi}\calK f\to\Pi_\calD\calK f$ with probability one as $N\to\infty$.
    \item[ii)] Consequently, for any fixed $k\in\N_{>0}$, the iterates $\{\phi^{\theta_1^{(k)}},\ldots,\phi^{\theta_m^{(k)}}\}$ of Algorithm~\ref{algo:data-driven} converge with probability one to the iterates $\{f_1^{(k)},\ldots,f_m^{(k)}\}$ of Algorithm~\ref{algo:idealized} (with the same $\calD$), as $N\to\infty$.
\end{itemize}
\end{corollary}

\begin{remark}[Convergence and approximation limits]\label{rem:three-limits}
It is useful to distinguish the different limiting arguments underlying Algorithm~\ref{algo:data-driven}.
(i) At the population level, Algorithm~\ref{algo:idealized} is a classical subspace iteration for the projected Koopman operator $K_\calD$.
Under the spectral-gap assumption $|\lambda_m|>|\lambda_{m+1}|$, its iterates satisfy
\[
    \spann\{f_1^{(k)},\ldots,f_m^{(k)}\}
    \longrightarrow \calV_m
    \qquad\text{as } k\to\infty,
\]
for almost every initialization.
(ii) Algorithm~\ref{algo:data-driven} replaces each exact projection appearing in this ideal iteration by an empirical projection computed from $N$ samples.
Corollary~\ref{cor:dd-convergence} shows that, for any fixed number of iterations $k$, this approximation becomes exact as $N\to\infty$: the data-driven iterates converge almost surely to the corresponding ideal iterates.
(iii) Finally, the neural implementation introduces an additional approximation, since the empirical projection is obtained by training a neural network.
The NTK argument below shows that, in the infinite-width regime, this training procedure can be interpreted as a linear empirical projection, thereby connecting the neural implementation to the setting of Corollary~\ref{cor:dd-convergence}.
\end{remark}

Finally, we discuss the assumption that $\Pi'_{N,\Phi}$ is a linear operator, made in Corollary~\ref{cor:dd-convergence} (and Remark~\ref{rem:ortho-nn}).
Note that it is in general not a linear operator since $\{\phi^\theta:\theta\in\R^{n_\theta}\}$ is not a linear subspace.\footnote{For instance, it is not guaranteed in general that for all $\theta_1,\theta_2\in\R^{n_\theta}$, there is $\theta\in\R^{n_\theta}$ such that $\phi^\theta=\phi^{\theta_1}+\phi^{\theta_2}$.}
Nevertheless, the \emph{neural tangent kernel} (NTK) theorem~\cite{jacot2018neural,golikov2022neural} allows us to assume that when the width of $\phi^\theta$ is large, $\Pi'_{N,\Phi}$ yields a linear operator.

\begin{theorem}[{\cite[Th.~2]{jacot2018neural}}]
Consider a neural network $\phi^\theta$ with random Gaussian parameter $\theta\in\R^{n_\theta}$.
In the infinite-width limit, i.e., when the hidden widths $c_1,\ldots,c_L\to\infty$, the neural tangent kernel
\[
\Theta_\theta(x,y) \triangleq \nabla_\theta \phi^\theta(x)^\top \nabla_\theta \phi^\theta(y), \quad x,y\in\calX,
\]
converges in probability (with respect to $\theta$) to a \emph{deterministic}\footnote{$\Theta_*$ depends only on the depth $L$, the activation function and the variance of the parameter distribution.} kernel $\Theta_*$.
Moreover, when trained by gradient flow on the $\ell^2$ loss (and with parameter $\theta$ initialized randomly as above), the network output evolves according to
\[
\frac{\mathrm{d}}{\mathrm{d}t} \phi^{\theta_t}(x) \propto -\sum_{i=1}^N \Theta_*(x,x_i)\big\{\phi^{\theta_t}(x_i)-f(x_i)\big\}.
\]
Hence, in the infinite-width regime, the training dynamics corresponds to \emph{kernel regression} in the reproducing kernel Hilbert space associated with $\Theta_*$.
\end{theorem}

The NTK theorem implies that, in the infinite-width regime and under gradient-flow training from the stated random initialization, the network output evolves in the linear span of the kernel sections $\Theta_*(\cdot,x_i)$ evaluated at the training inputs.
Consequently, least-squares training acts as kernel regression in the finite-dimensional data-dependent space
\[
\calD_N=\spann\{\Theta_*(\cdot,x_i)\}_{i=1}^N.
\]
In this asymptotic regime, the fitted-value map $f\mapsto\Pi'_{N,\Phi}f$ can therefore be identified with the empirical linear projection $\Pi'_{N,\calD_N}$ (up to the usual qualifications associated with initialization and optimization in the NTK limit).
This is the sense in which the nonlinear neural parametrization induces the linear projection assumed in Corollary~\ref{cor:dd-convergence}; it is not an assumption that finite-width neural networks themselves form a linear function class.
Thus, Corollary~\ref{cor:dd-convergence} provides an asymptotic justification for the neural implementation.\footnote{We omit the verification that $\calD\subseteq\calF$ and $\calK\calD\subseteq\calF$ because this is expected to hold generically since neural networks are piecewise smooth.}

%%%%%%%%%%%%%%%%%%%%%%%%%%%%%%%%%%%%%%%%%%%%%%%%%%%%%%%%%%%%%%%%%%%%%%%%%%%%%%%%
\section{Numerical experiments}\label{sec:experiments}

\ifextended

In all experiments, we used neural networks with $L=2$ hidden layers, with widths denoted by $c_1$ and $c_2$, and activation function $\sigma=\tanh$.
Using PyTorch~\cite{paszke2019pytorch}, we trained the networks over $800$ epochs with AdamW optimizer with learning rate $10^{-3}$, no weight decay, and full-batch training.
The maximum number of iterations of Algorithm~\ref{algo:data-driven} was $k_{\mathrm{max}}=50$.
This fixed budget was used uniformly in the reported experiments; more generally, one may stop when successive empirical subspaces (or the associated Koopman residuals) vary below a prescribed tolerance.
We compared with extended dynamic mode decomposition (EDMD)~\cite{williams2015adatadriven}.

Code for the implementation of the method and the numerical experiments is available at \url{https://github.com/guberger/DeepKoopmanLearning}.
All computations were made on a laptop with processor Intel Core i7-7600u and 16GB RAM running Windows.

\subsection{Nonlinear 1D system}

Consider the system on $\calX=\R$ given by
\begin{equation}\label{eq:nonlinear-1d}
x_{t+1} = 0.9 \tanh(x_t).
\end{equation}
This system admits a globally asymptotically stable equilibrium point at $x=0$.

We applied Algorithm~\ref{algo:data-driven} on this system with $m=3$.
We used $N=2500$ samples and standard Gaussian distribution.
We used hidden widths $c_1=c_2=8$.

We denote by $g_1,\ldots,g_m$ the computed approximations of the $m$ dominant Koopman modes of the system, normalized so that $\lVert g_i\rVert'_N=1$.
Those are represented in Figure~\ref{fig:tanh-1d-nn}.
We have also represented $\lambda_i^{-1}\calK g_i^{}$, where $\lambda_i$ are the associated eigenvalues.
Good approximations satisfy $\lVert\lambda_i^{-1}\calK g_i^{} - g_i^{}\rVert$ is small~\cite{colbrook2024rigorous}.
The values are given in Table~\ref{table:error-tanh-1d-nn}.
We see that the approximation error is overall small ($\leq 0.05$).

\begin{figure}[h]
    \centering
    \includegraphics[width=\linewidth]{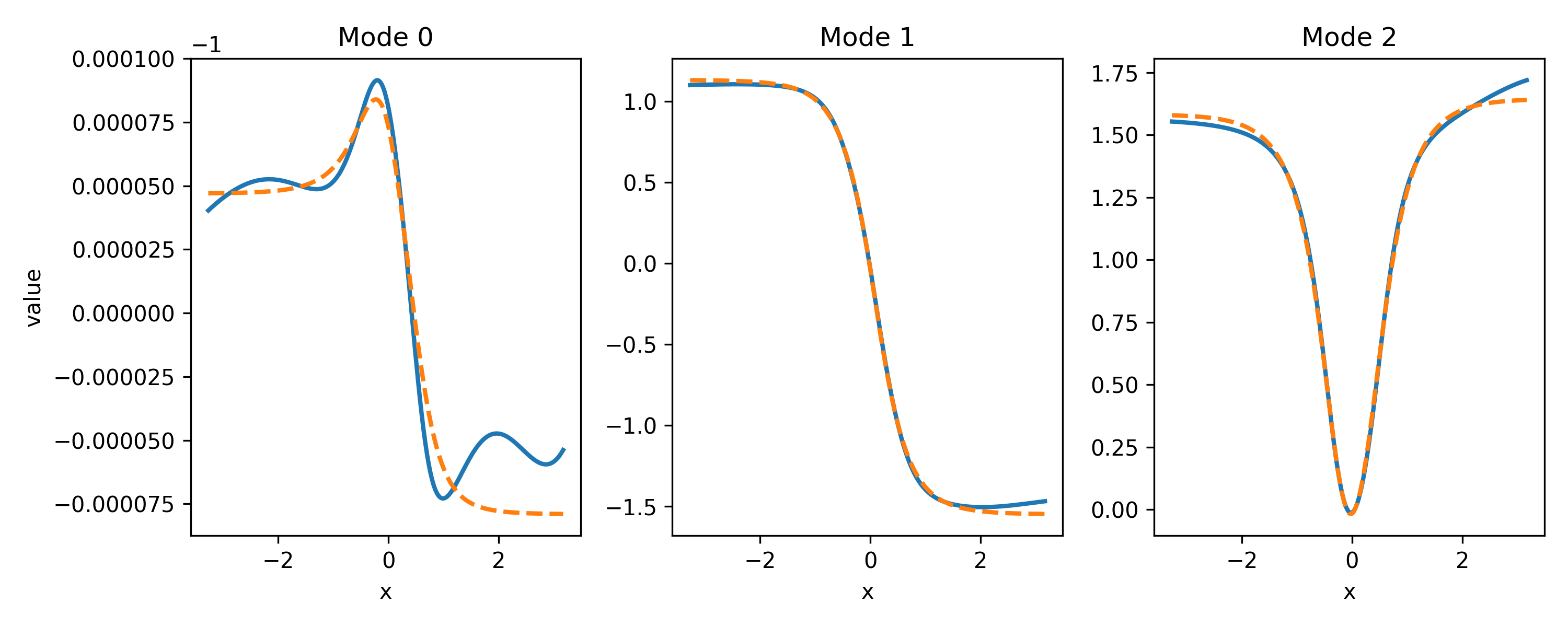}
    \caption{Approximation of dominant $m=3$ Koopman modes of system~\eqref{eq:nonlinear-1d} with Algorithm~\ref{algo:data-driven}.
    The blue lines represent the normalized Koopman modes $g_i$.
    The orange lines represent $\lambda_i^{-1}\calK g_i^{}$.
    As we can see, $\lambda_i^{-1}\calK g_i^{}\approx g_i$.
    Note that the shifted value axis for $g_0$.}
    \label{fig:tanh-1d-nn}
\end{figure}

\begin{table}[h]
    \centering
    \begin{tabular}{@{}cccc@{}}
    \toprule
    $i$ & $1$ & $2$ & $3$ \\
    \midrule
    $\lVert\lambda_i^{-1}\calK g_i^{} - g_i^{}\rVert'_N$ & $1.1\cdot10^{-5}$ & $1.2\cdot10^{-2}$ & $1.4\cdot10^{-2}$ \\
    \bottomrule
    \end{tabular}
    \caption{Approximation error of Algorithm~\ref{algo:data-driven} on system~\eqref{eq:nonlinear-1d}.}
    \label{table:error-tanh-1d-nn}
    \vspace{-5mm}
\end{table}

Next, we compared with EDMD with $\calD$ being the set of polynomials of degree up to $3$ or $6$.\footnote{We note that for degrees $>6$ numerical errors and spectral pollution caused the approximation to be actually \emph{worse}.}
The results are presented in Figure~\ref{fig:tanh-1d-poly} and Table~\ref{table:error-tanh-1d-poly}.
Overall, one sees that our approach provides smaller approximation errors.
The computation time is however longer: $30$ secs for our approach vs.~$2$ secs for the EDMD approach.
This additional cost is expected because each power iteration requires fitting several neural networks, whereas EDMD solves a single finite-dimensional regression problem.

\begin{figure}[h]
    \centering
    \includegraphics[width=\linewidth]{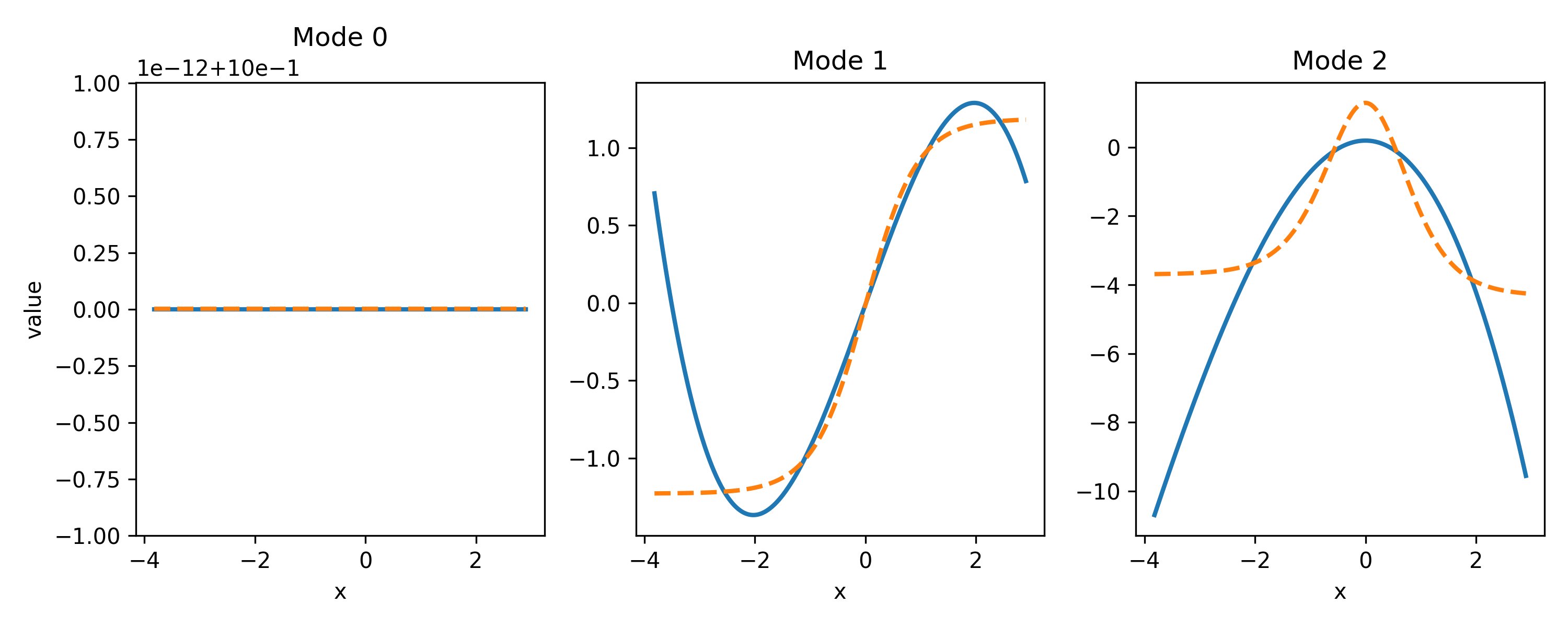}\\
    \includegraphics[width=\linewidth]{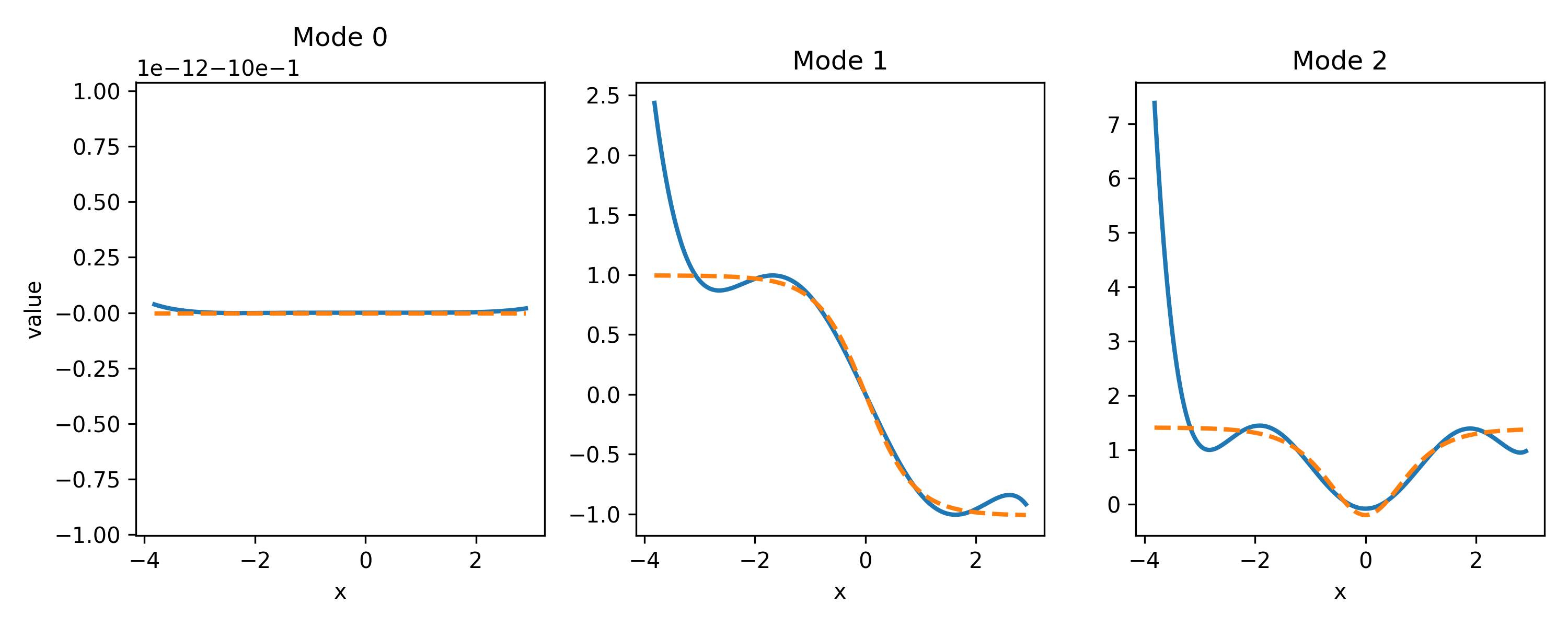}
    \caption{Approximation of the dominant three Koopman modes of system~\eqref{eq:nonlinear-1d} with EDMD using polynomials of degree up to $3$ (top) or $6$ (bottom).
    The color code is the same as in Figure~\ref{fig:tanh-1d-nn}.
    Note that the shifted value axis for $g_0$.}
    \label{fig:tanh-1d-poly}
\end{figure}

\begin{table}[h]
    \centering
    \begin{tabular}{@{}cccc@{}}
    \toprule
    $i$ & $1$ & $2$ & $3$ \\
    \midrule
    \multirow{2}{*}{$\lVert\lambda_i^{-1}\calK g_i^{} - g_i^{}\rVert'_N$} & $2.4\cdot10^{-15}$ & $1.2\cdot10^{-1}$ & $6.0\cdot10^{-1}$ \\
    & $3.7\cdot10^{-15}$ & $7.4\cdot10^{-2}$ & $2.1\cdot10^{-1}$ \\
    \bottomrule
    \end{tabular}
    \caption{Approximation error of EDMD with polynomials of degree up to $3$ (1\textsuperscript{st} row) or $6$ (2\textsuperscript{nd} row) on system~\eqref{eq:nonlinear-1d}.}
    \label{table:error-tanh-1d-poly}
    \vspace{-5mm}
\end{table}

\subsection{Bistable system}

Next, consider the continuous-time system on $\calX=\R^2$ described by the unforced Duffing equation:
\begin{equation}\label{eq:duffing}
\left\{\begin{array}{rcl}
    \dot{x}_1^{}(t) &=& x_2^{}(t), \\
    \dot{x}_2^{}(t) &=& -0.5x_2^{}(t) + x_1^{}(t) - x_1^3(t).
\end{array}\right.
\end{equation}
This system has two asymptotically stable equilibrium points at $x_{*,1}=(-1,0)$ and $x_{*,2}=(1,0)$, and one saddle point at $x_{*,3}=(0,0)$ (see Figure~\ref{fig:duffing-basins}).
Using our approach, we will estimate the basins of attraction of $x_{*,1}$ and $x_{*,2}$, by computing nontrivial dominant eigenfunctions of the Koopman operator of the associated discrete-time system.
We stress out that obtaining good approximations is a nontrivial problem~\cite{takeishi2017learning} since this requires discontinuous approximation functions.

\begin{figure}[h]
    \centering
    \includegraphics[width=0.8\linewidth]{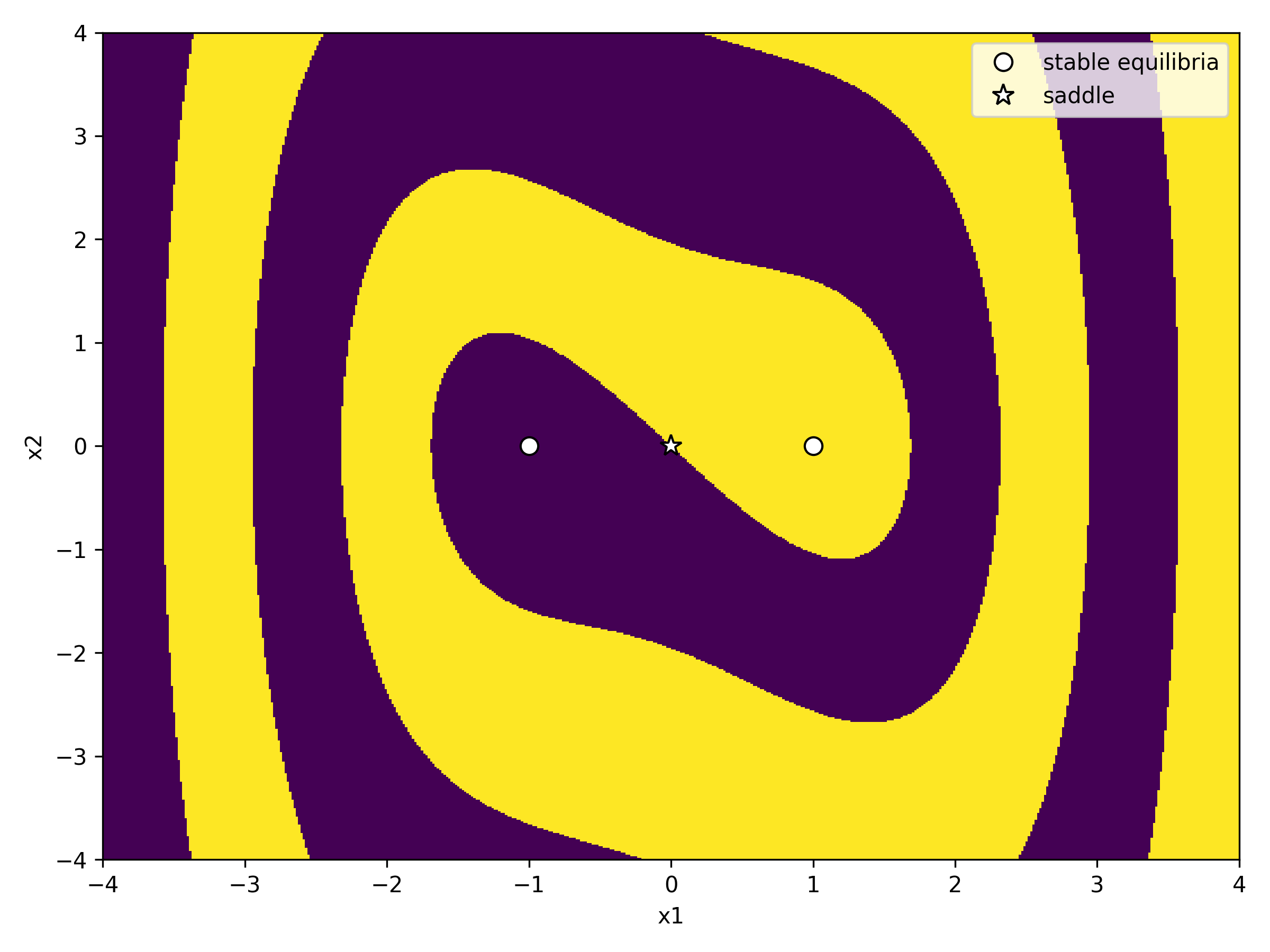}
    \caption{Basins of attraction of Duffing system~\eqref{eq:duffing}, estimated using numerical simulations.}
    \label{fig:duffing-basins}
\end{figure}

Concretely, we first time-discretized the system with sampling period $T=1$.\footnote{This means that $F(x)=\xi(1)$ where $\xi(\cdot)$ is the trajectory of \eqref{eq:duffing} with $\xi(0)=x$.
Numerically, this was computed using fourth-order Runge--Kutta with time step $dt=0.01$.}
Then, we applied Algorithm~\ref{algo:data-driven} on the discrete-time system with $m=2$.
We used $N=10^4$ samples and uniform distribution on $[-4,4]^2$.
We used hidden widths $c_1=c_2=64$.

We denote by $g_1,\ldots,g_m$ the computed approximations of the $m$ dominant Koopman modes of the system, normalized so that $\lVert g_i\rVert'_N=1$.
Those are represented in Figure~\ref{fig:duffing-nn}.
We see that the flat components of the modes $g_i$ qualitatively recover the basins of attraction of the system.
This experiment is intended as an illustration of the structure encoded by the learned eigenfunctions; it does not constitute a certified region-of-attraction computation.
We have also represented $\lambda_i^{-1}\calK g_i^{}$, where $\lambda_i$ are the associated eigenvalues.
The values of $\lVert\lambda_i^{-1}\calK g_i^{} - g_i^{}\rVert$ are given in Table~\ref{table:error-duffing-nn}.
We see that the approximation error is overall small ($\leq0.1$).
The computation time was approx.~$8$ min.

\begin{figure}[h]
    \centering
    \includegraphics[width=\linewidth]{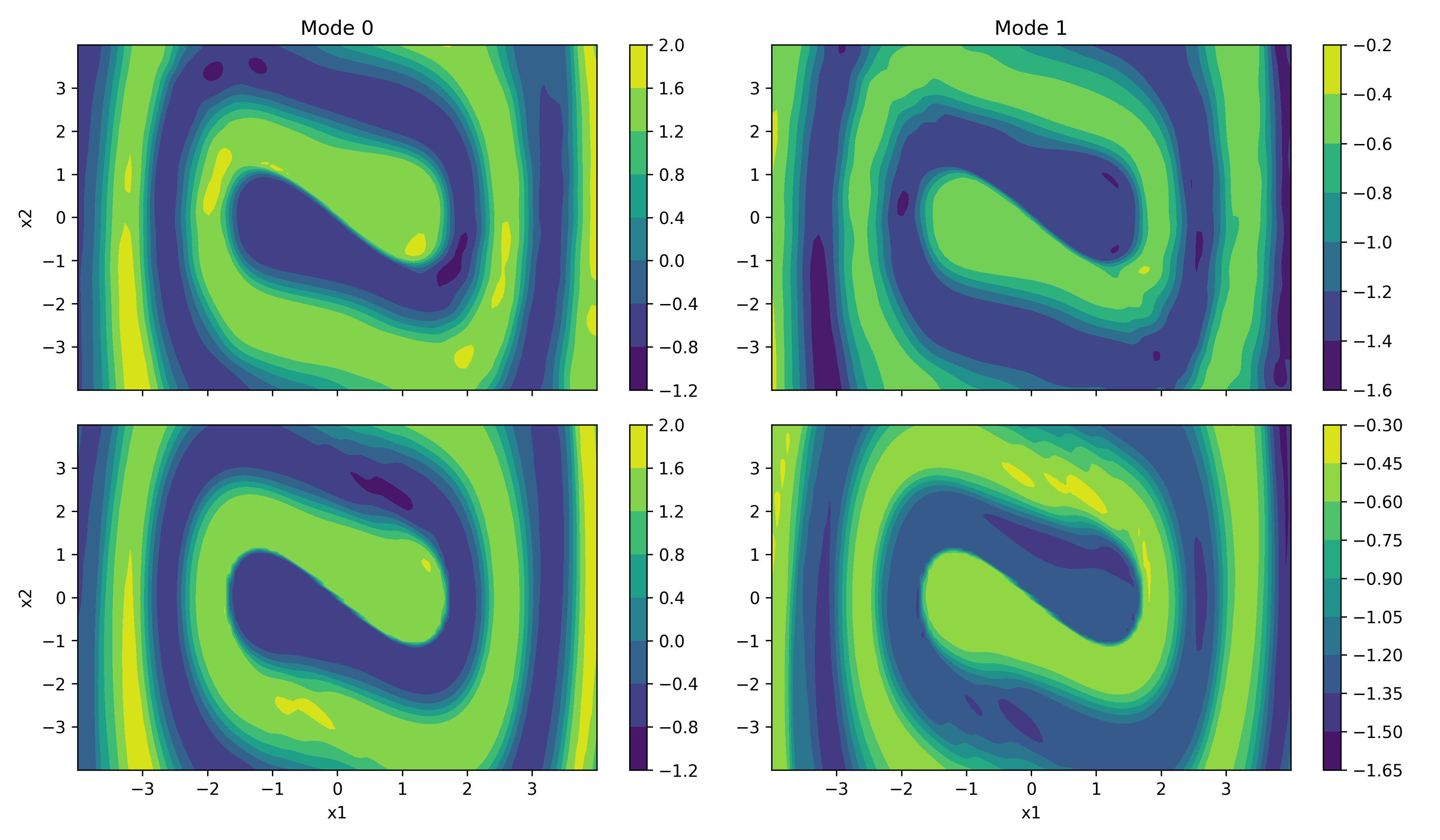}
    \caption{Approximation of dominant $m=2$ Koopman modes of system~\eqref{eq:duffing} with Algorithm~\ref{algo:data-driven}.
    The contour plots on the top represent the normalized Koopman modes $g_i$.
    The contour plots on the bottom represent $\lambda_i^{-1}\calK g_i^{}$.
    As we can see, $\lambda_i^{-1}\calK g_i^{}\approx g_i$.}
    \label{fig:duffing-nn}
\end{figure}

\begin{table}[h]
    \centering
    \begin{tabular}{@{}cccc@{}}
    \toprule
    $i$ & $1$ & $2$ \\
    \midrule
    $\lVert\lambda_i^{-1}\calK g_i^{} - g_i^{}\rVert'_N$ & $0.10$ & $0.057$ \\
    \bottomrule
    \end{tabular}
    \caption{Approximation error of Algorithm~\ref{algo:data-driven} on system~\eqref{eq:duffing}.}
    \label{table:error-duffing-nn}
    \vspace{-5mm}
\end{table}

Next, we compared with EDMD with $\calD$ being the set of polynomials of degree up to $3$ or $6$.\footnote{We note that for degrees $>6$ numerical errors and spectral pollution caused the approximation to be actually \emph{worse}.}
The results are presented in Figure~\ref{fig:duffing-poly} and Table~\ref{table:error-duffing-poly}.
We see that this approach fails to provide good approximations of the dominant Koopman modes (Table~\ref{table:error-duffing-poly}) and of the basins of attraction of the system (Figure~\ref{fig:duffing-poly}).
The main reason for that is the inability to represent discontinuous functions.

\begin{figure}[h]
    \centering
    \includegraphics[width=\linewidth,trim=0cm 9.0cm 0cm 0cm,clip]{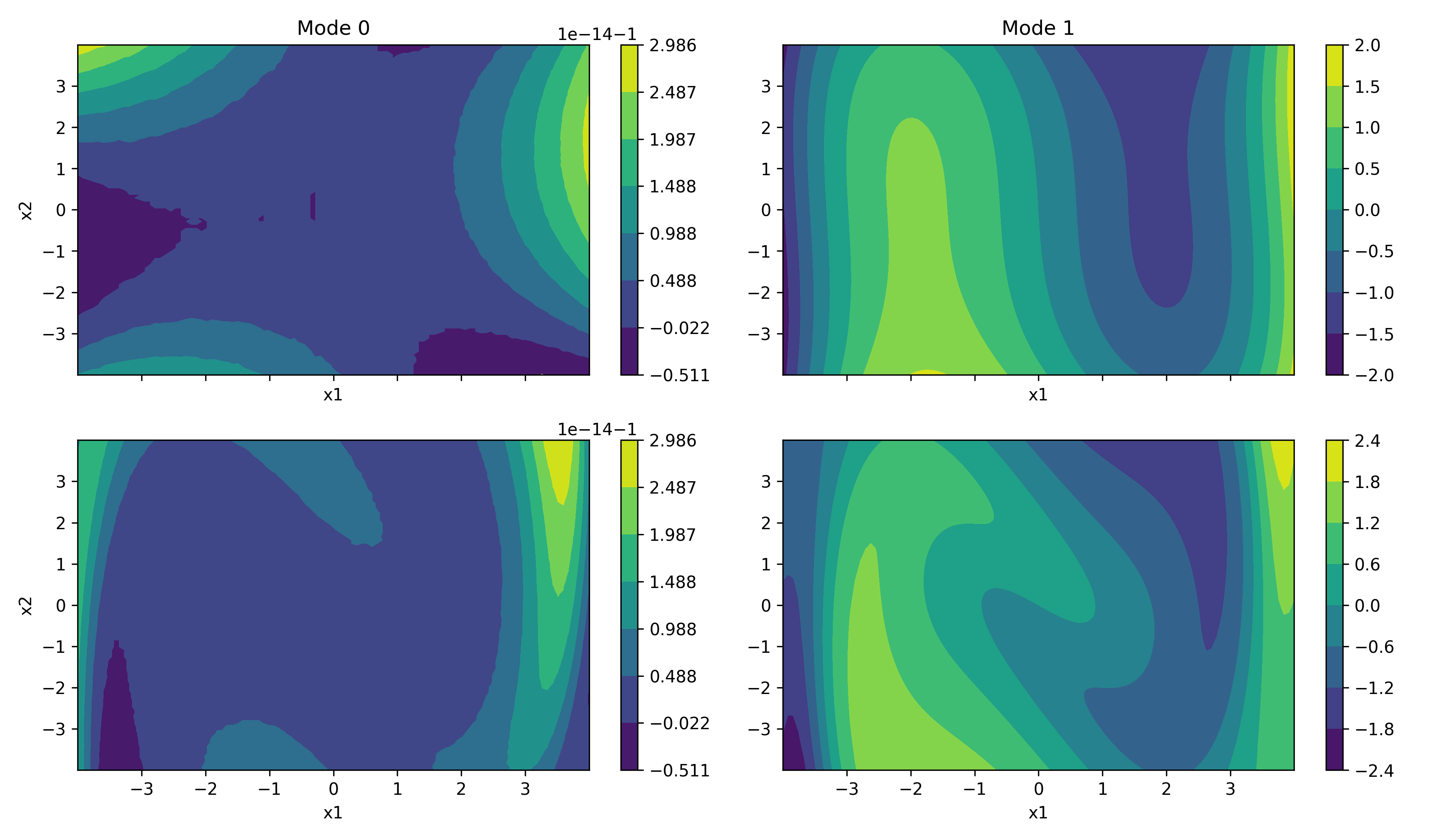}\\
    \includegraphics[width=\linewidth,trim=0cm 9.0cm 0cm 0cm,clip]{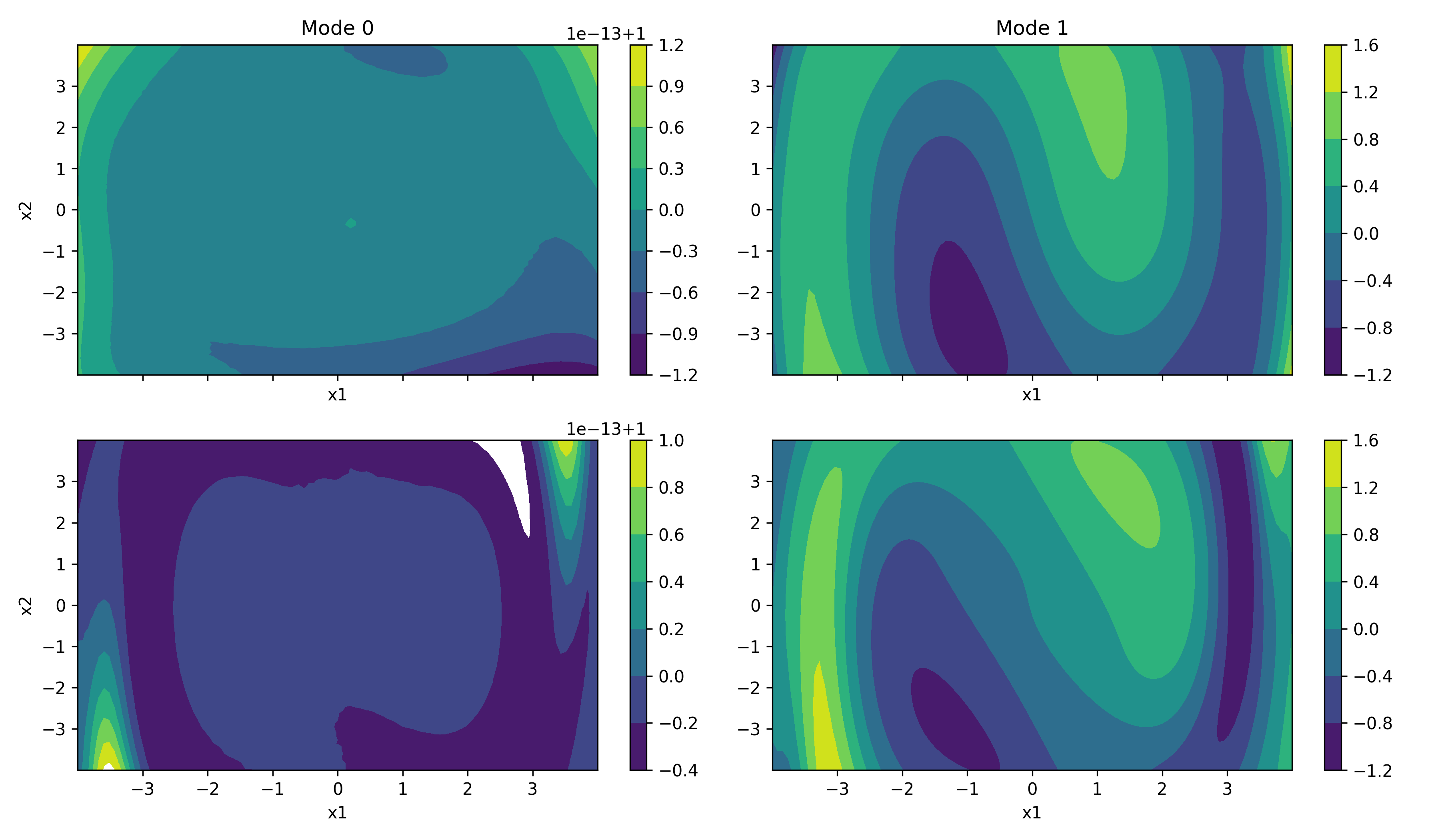}
    \caption{Approximation of the dominant three Koopman modes of system~\eqref{eq:duffing} with EDMD using polynomials of degree up to $3$ (top) or $6$ (bottom).
    The contour plots represent the normalized Koopman modes $g_i$.
    Note that the shifted value axis for $g_0$.}
    \label{fig:duffing-poly}
\end{figure}

\begin{table}[h]
    \centering
    \begin{tabular}{@{}cccc@{}}
    \toprule
    $i$ & $1$ & $2$ \\
    \midrule
    \multirow{2}{*}{$\lVert\lambda_i^{-1}\calK g_i^{} - g_i^{}\rVert'_N$} & $4.6\cdot10^{-14}$ & $5.8\cdot10^{-1}$ \\
    & $2.0\cdot10^{-14}$ & $5.8\cdot10^{-1}$ \\
    \bottomrule
    \end{tabular}
    \caption{Approximation error of EDMD with polynomials of degree up to $3$ (1\textsuperscript{st} row) or $6$ (2\textsuperscript{nd} row) on system~\eqref{eq:duffing}.}
    \label{table:error-duffing-poly}
    \vspace{-5mm}
\end{table}

\else

See extended version~\cite{berger2026datadriven}.

\fi

%%%%%%%%%%%%%%%%%%%%%%%%%%%%%%%%%%%%%%%%%%%%%%%%%%%%%%%%%%%%%%%%%%%%%%%%%%%%%%%%
\section{Scope and limitations}

\ifextended

The method targets dominant spectral components of the projected operator $K_\calD$ and therefore inherits the usual limitations of finite-dimensional Koopman approximations, including possible spectral pollution and dependence on the chosen approximation space.
The number of requested modes $m$ is a user parameter: increasing $m$ targets a larger invariant subspace but requires additional regressions and, for convergence of subspace iteration, an appropriate separation between the retained and discarded parts of the spectrum.
Moreover, the consistency result above is asymptotic and does not provide a finite-sample error bound or a universal prescription for the number of data points; the required data depend on the dynamics, sampling distribution, approximation class, and spectral/conditioning properties of the problem.
The NTK argument is likewise an infinite-width justification, while the experiments necessarily use finite networks and numerical optimization.
Finally, the repeated neural regressions make the proposed method computationally more expensive than EDMD in the examples considered here.
Reducing this computational burden, deriving finite-sample guarantees, and studying principled choices of $m$ and stopping criteria are important directions for future work.

\else

See extended version~\cite{berger2026datadriven}.

\fi

%%%%%%%%%%%%%%%%%%%%%%%%%%%%%%%%%%%%%%%%%%%%%%%%%%%%%%%%%%%%%%%%%%%%%%%%%%%%%%%%
\section{Conclusions}

\ifextended

This work introduced a novel data-driven algorithm to approximate dominant eigenfunctions of a projected Koopman operator using a neural-network-based power iteration scheme.
By avoiding the explicit construction of the projected Koopman operator and bypassing the need for predefined dictionaries or anti-collapse regularization mechanisms, the proposed method addresses limitations of existing approaches.

The theoretical analysis separates the classical convergence of the ideal subspace iteration in the iteration index from the consistency of its data-driven approximation for increasing sample size, while the NTK limit provides an asymptotic justification of the linear-projection interpretation of neural regression.
This provides a principled justification for the method, linking it to classical operator theory and modern neural tangent kernel results.
Empirical results demonstrate that the approach accurately captures dominant Koopman modes, including challenging cases such as bistable systems where eigenfunctions exhibit discontinuities.
In particular, the method successfully recovers meaningful structures such as basins of attraction, which are known to be encoded in eigenfunctions associated with eigenvalue one.
Compared to traditional techniques like EDMD with polynomials, the proposed method achieves improved accuracy, especially when expressive representations are required.

Overall, this work highlights the effectiveness of combining operator-theoretic ideas with neural networks in a fully data-driven framework.
Future directions include extending the approach to controlled systems, improving computational efficiency, and further investigating robustness to noise and finite-sample effects.

\else

See extended version~\cite{berger2026datadriven}.

\fi

%%%%%%%%%%%%%%%%%%%%%%%%%%%%%%%%%%%%%%%%%%%%%%%%%%%%%%%%%%%%%%%%%%%%%%%%%%%%%%%%

\bibliographystyle{IEEEtran}
\bibliography{myrefs}

\end{document}